# Quantum-assisted and Quantum-based Solutions in Wireless Systems

Sandor Imre, *Member, IEEE* and Laszlo Gyongyosi, *Member, IEEE,*

Department of Telecommunications

Budapest University of Technology and Economics

Budapest, Hungary

*Abstract*— **In wireless systems there is always a trade-off between reducing the transmit power and mitigating the resultant signal-degradation imposed by the transmit-power reduction with the aid of sophisticated receiver algorithms, when considering the total energy consumption. Quantum-assisted wireless communications exploits the extra computing power offered by quantum mechanics based architectures. This paper summarizes some recent results in quantum computing and the corresponding application areas in wireless communications.**

## I. INTRODUCTION

One day in 1965 when Gordon Moore from Intel was preparing a presentation and started to draw a plot about the performance of memory chips he suddenly observed an interesting rule-of-thumb, which was later termed as Moore's law. He concluded that since the invention of the transistor the number of transistors per chip roughly doubled every 18-24 months, which eventually resulted in an exponential increase in the computing power of computers. Although this was an empirical observation without any theoretical substantiation, Moore's law seems to have maintained its validity over the years, provided of course that sufficient investment in science and technology is attracted by the semiconductor industry.

The growth in the processors' performance is due to the fact that we put more and more transistors on the same size chip. This requires smaller and smaller transistors, which can be achieved if we are able to draw thinner and thinner lines onto the surface of a semiconductor wafer, lines that are significantly thinner than hair. Current semiconductor technology also enables us to remove or retain certain parts of the wafer according to the specific layout of transistors, diodes, external pins, etc.

If the current trend of miniaturization continues, above-mentioned lines will depart from the well-known natural environment obeying the well-understood rules revealed step by step during the evolution of the human race and enter into a new world, where 'the traveler has to obey strange new rules if he/she would like to pass through this nano-world'. The new rules are explained by quantum mechanics and the 'border between these two worlds' lies around one nanometer ($10^{-9}$m) thickness. These rules are sometimes similar to their classic (i.e. macroscopic) counterparts, but sometimes they are quite strange. The reality is though that we have entered this 'nano-era', hence we have to accept its rules as the new framework of computing and communications. Let us briefly explore their benefits.

### A. Background

In the year 1985 Feynman suggested a new straightforward approach [138]. Instead of regarding computers as devices operating under the laws of classic physics - which is common sense – let us consider their operation as a special case of a more general theory governed by quantum mechanics [139]. Our goal is that of seeking algorithms, which are more efficient than their best classic counterparts, but are only available in the quantum world. The corresponding software-related efforts are in the realms of quantum computing [140]. We might also hypothesize that the capacity of a quantum channel could exceed that of classic wireless links or that we could design more secure protocols than the currently applied ones. Quantum communications [141] or quantum information theory [142] aims for answering these questions.

In order to understand how quantum computing and communication might improve the performance of our classic wireless systems, let us summarize the four basic rules (called Postulates) of quantum mechanics from a telecommunications engineering point of view [140]. These are similar to the Euclidean axioms of geometry in the sense that there are no formal proofs supporting them - but as a difference - anyone who presents an experiment, which contradicts to the postulates might stand a chance of receiving the Nobel Prize. We will demonstrate that any reader who is well-versed in wireless communications has sufficient background to accept these rules at an abstraction level, which is required to absorb the results presented in this paper. In order to pave the way further, we will invoke the well-known DS-CDMA Maximum Likelihood Multi User Detection (ML-MUD) example as a bridge between the classic wireless and the quantum world.

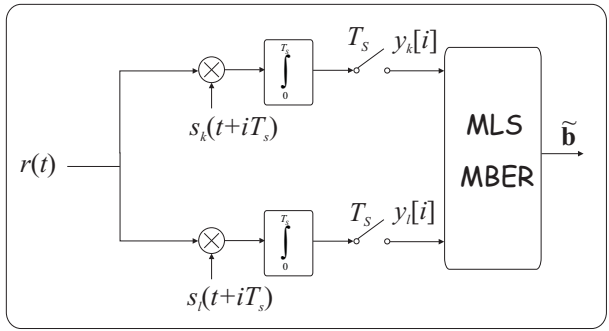

Fig. 19: Multi-user DS-CDMA detector

In our fairly simplified model the $i^{th}$ symbol of the $k^{th}$ ($k = 1, 2, \ldots, K$) user is denoted by $b_k[i]$, where the symbol-duration is $T_S$. For the sake of simplicity we opt for BPSK, i.e. we have $b_k[i] \in \{+1, -1\}$. The channel-induced signal distortion from the $k^{th}$ user's perspective is modeled by means of $h_k(i, t) = a_k[i]\delta(t - \tau_k)$, where we have $a_k[i] = A_k[i]e^{j\alpha_k[i]}$ and $A_k[i], \alpha_k[i]$ as well as $\tau_k$ are typically independent random variables. Finally, $s_k(t)$ refers to the unique, user-specific DS-CDMA signature waveform.

The complex baseband equivalent representation of the signal received at the base station is calculated by convolving the channel's input with its impulse response in the following manner:

$$r(t) = \sum_{k=1}^{K} h_k(i, t) * v_k(i, t) = \sum_{k=1}^{K} a_k[i]b_k[i]s_k(t - iT_S - \tau_k). \tag{13}$$

Since in the uplink different $\tau_k$ delays are considered owing to the different distances of the MSs from the BS, the system is *asynchronous*. Furthermore, $a_k[i]$ is assumed to be completely unknown at the receiver, hence we have to solve a *blind* MUD problem.

When applying matched filters (MF) in the BS's detector, their output for the $i^{th}$ symbol may be denoted by $y_k[i]$

$$y_k[i] = \int_{iT_s}^{(i+1)T_s} r(t)s_k(t - iT_s)dt. \tag{14}$$

In order to formulate the detection Cost Function (CF) more explicitly, we construct two matrices from the transmitted symbol combinations and from the corresponding MF outputs, yielding $\mathbf{b} = [b_1, b_2, \ldots, b_K]$, $\mathbf{y} = [y_1, y_2, \ldots, y_K]$, respectively.

To expound further, given the received signals encapsulated in the MF output vector $\mathbf{y}$, we have $2^K$ different hypotheses according to all the different legitimate transmitted signals hosted by the vectors $\mathbf{b}_m$, yielding the vector $\mathbf{y} = w(\mathbf{b}_m)$ hosting all the received signals. The vector-to-vector mapping function $w(\cdot)$ represents the matched filters' outputs in response to the transmitted symbol-vector $\mathbf{b}_m$ containing the symbols transmitted by all the users. More explicitly, this represents the $m^{th}$ hypothesis. The corresponding MUD architecture is depicted in Fig. 19.

Obviously $w(\cdot)$ depends not only on the transmitted symbols of the $K$ users, but also on the random channel parameters. Moreover the mapping $w(\cdot)$ is not reversible. Therefore we are unable to unambiguously identify that particular transmitted symbol vector $\mathbf{b}$, which results exactly in the received symbol vector $\mathbf{y}$. Instead, the optimal decision relying on the Maximum Likelihood Sequence (MLS) CF 'simply' requires us to spot that particular hypothesis

$m$, which maximizes the conditional probability density function (PDF) of

$$\tilde{\mathbf{b}}_{MLS} : \max_m f(\mathbf{y}|\mathbf{b}_m). \tag{15}$$

**1.** The $1^{st}$ Postulate declares how to describe the state of any physical system [14].

Considering our ML MUD problem, similarly to the classic vectors $\mathbf{b}$ or $\mathbf{y}$, a quantum register consisting of $K$ *qbits* stores $2^K$ legitimate states at any instant, but the quantum register may assume all these states simultaneously, i.e. in parallel, which is formulated as [141]:

$$|\varphi\rangle = \sum_{i=0}^{2^K - 1} \varphi_i |i\rangle. \tag{16}$$

This implies from our MUD perspective that a single quantum register is capable of storing all the legitimate $\mathbf{b}_m$ hypotheses.

**2.** The $2^{nd}$ Postulate is related to the time evolution of any system in time domain. [15]

The parallel processing capability of the quantum-search originating from the $2^{nd}$ Postulate allows us to find the discrete PDF given by the relative frequencies of those transmitted signal vectors $\mathbf{b}_m$ that lead to a certain received signal vector $\mathbf{y}$, which are then combined by weighting with the corresponding *a priori* probabilities. Finally, in possession of the function $f(\cdot|\cdot)$ we may opt for using quantum-search for finding the most likely transmitted signal vector $\mathbf{b}_m$, given a specific received signal vector $\mathbf{y}$.

This postulate is one of the key features responsible for the significant speed up of quantum algorithms. Similar to the classic algorithms, any quantum-domain algorithm - such as a unitary transform - can be decomposed into a set of two-and four dimensional unitary transforms (like the Karnaugh method in classical systems) and implemented by means of a predefined set of corresponding elementary quantum gates. Returning to our ML MUD example - as any classic detector, this ML MUD may be implemented using adders, multipliers, inverters, logical NAND etc. By analogy, provided that we can find an efficient quantum MUD algorithm, we will be able to implement it using basic quantum gates and circuits. This analogy might appear to be trivial in the light of our everyday

---

[14] In quantum computing, a qbit or quantum bit is a unit of quantum information, namely the 'quantum' counterpart of the classic bit. More explicitly, the qbit is described by a specific quantum state in a two-state quantum-mechanical system, which is formally equivalent to a two-dimensional vector space defined over the complex numbers. A specific example of a two-state quantum system is constituted by the two legitimate polarizations of a single photon, namely the vertical and horizontal polarizations. In a classic system, a bit would have to be either a logical 'one' or a logical 'zero', but apart from 'one' and 'zero' quantum mechanics allows the qbit to be concomitantly in a superposition of both states, which inherently facilitates their parallel processing. This beneficial property is inherent in quantum computing.

To elaborate a little further, a single qbit may be represented as $|\varphi\rangle = a|0\rangle + b|1\rangle$, where $|\rangle$ is referred to as the Dirac's Ket-notation [143] routinely used in quantum-physics for describing a state, while $a$ and $b$ are complex-valued numbers satisfying $|a|^2 + |b|^2 = 1$. Hence the qbit may be interpreted as a vector in the two-dimensional complex-valued vector-space, where $a$ and $b$ are the complex-valued probability amplitudes within the orthogonal bases $|0\rangle$ and $|1\rangle$ of the vector-space $|\varphi\rangle$.

[15] From an engineering point of view the Schrödinger equation simplifies to the following essence: the evolution of any closed physical system may be characterized with the aid of unitary transforms obeying the property of $U^{-1} = U^{\dagger}$, and $|\psi\rangle = U|\varphi\rangle$, where $U^{\dagger}$ denotes the complex conjugated and transposed version of $U$.

practice, but owing to the strange rules of quantum mechanics it is not trivial at all - quite the contrary, it is remarkable.

3. The $3^{rd}$ Postulate connects the nano as well as the classic macroscopic world and it is referred to as 'the measurement'.[16]

   From a ML MUD's perspective the received signal vector $\mathbf{y}$ is entered into the quantum detector, which also prepares an additional register containing all legitimate hypotheses $\mathbf{b}_m$ associated with uniformly distributed coefficients according to the $1^{st}$ Postulate. Hence, if we performed a measurement on this register, we would find that any of the hypotheses $\mathbf{b}_m$ has a probability of $1/2^K$. The $2^{nd}$ Postulate enables us to modify these coefficients in such a way that the most likely hypothesis will have the largest coefficient. An appropriately constructed measurement will deliver this particular hypothesis with a probability that is proportional to the absolute squared value of the largest coefficient. In order to increase this probability towards unity, we have to conceive a sophisticated quantum algorithm and prepare an appropriate measurement.

4. The $4^{th}$ Postulate defines the technique of combining individual quantum systems. In our ML MUD example the quantum register containing the hypotheses $\mathbf{b}_m$ has a length of $K$ qbits. It may however also be viewed as being constituted by $K$ independent qbits. This postulate sets out the rules of how to switch between the above-mentioned two different perspectives. Furthermore, if we would like to combine two detectors, each designed for $K$ users, this postulate outlines how to construct the register having a length of $2K$ in the resultant joint detector.[17]

In conclusion of the basic rules, we emphasize that classic physics and engineering may be regarded as a subset of quantum theory. More explicitly, the postulates discussed above extend the design-space of practical algorithms and protocols we may invoke in telecommunications problems. However, *this new quantum world simply opens up new realms of solutions, without actually telling us, how to construct these solutions. In this respect, its role is reminiscent of Shannonian information theory - it took our community over half-a-century to find near-capacity solutions capable of approaching Shannon's visionary predictions...*

---

[16] Any quantum measurement may be described by means of a set of measurement operators $\{M_m\}$, where $m$ stands for a legitimate classic integer result of the measurement. Quantum measurements differ in two aspects from classic ones. Firstly, they are random in the sense that getting the specific result $m$ has a certain probability. Secondly, the measurement itself typically influences/perturbs or modifies the measured object. The role of the measurement may be deemed to be analogous to that of the D/A converter in the classic 'A/D converter, Digital Signal Processing (DSP), D/A converter' chain. More explicitly, in quantum-processing we have a Classic-to-Quantum (C/Q) domain converter, followed by a quantum-algorithm and a Q/C converter, where again, the latter block carries out the measurement.

[17] Individual qbits have to be combined by means of the tensor product denoted as $\otimes$ exactly in the same manner as in case of classic bits (e.g. $0 \otimes 1 \Rightarrow 01$). When considering several qbits, such as for example a 2-qbit system, there are four quantum states i.e. $|00\rangle$, $|01\rangle$, $|10\rangle$ and $|11\rangle$, which are constituted by the tensor product of the $1^{st}$ and $2^{nd}$ qbits. The superimposed state is then formulated as

$$|\psi\rangle = a_{00}|00\rangle + a_{01}|01\rangle + a_{10}|10\rangle + a_{11}|11\rangle, \quad (17)$$

where $|a_{ij}|^2$ is the post-measurement probability of occurrence of the state $|ij\rangle$ normalized as $\sum_{i,j}|a_{ij}|^2 = 1$. Recall that a tensor product may be viewed as a bilinear operator, which is a function that linearly combines the elements of two vector spaces to generate an element in a third vector space, as exemplified by a matrix multiplication.

Before delving into any further discussions on quantum communication based solutions conceived for future wireless systems, one of the interesting but strange consequences of the above postulates should be mentioned:

- The *no cloning theorem* [141] of quantum computing claims that only known and/or orthogonal quantum states can be copied. Fortunately classic states, which are widely used in operational computers and processors are orthogonal, therefore *the quantum description of nature as a whole is in harmony with our every-day experiences.* This property is useful, if someone would like to protect his/her information against eavesdropping in the communication channel. In simplified terms we could argue that it is sufficient to encode the classic information states into non-orthogonal quantum states and as a result, the malicious hacker will become unable to make a copy. However, the *no cloning theorem* imposes a strict limitation, when constructing ML MUD detectors. For example, if we have an inner quantum state computed within the MUD algorithm, it is impossible to make several copies of it in order to perform different calculations on them! One of the most important current research challenges is that of designing repeaters for large wireless and optical quantum networks.

- *Entanglement [140]:* Let us now investigate further, what is encapsulated in the notation $|\varphi\rangle$. As argued in [140], when we consider a 2-qbit quantum state formulated as $|\varphi\rangle = a|00\rangle + b|11\rangle$, the question arises, whether it might be possible to decompose it into two individual single-qbit states, such as:

$$\begin{aligned}|\psi\rangle &= (a_1|0\rangle + b_1|1\rangle)(a_2|0\rangle + b_2|1\rangle) & (18)\\ &= a_1a_2|00\rangle + a_1b_2|01\rangle + b_1a_2|10\rangle + b_1b_2|11\rangle?\end{aligned}$$

The answer to this question may be shown to be *no*! To elaborate a little further, a latent linkage exists between the two qbits and hence if we decided to measure or determine the first qbit of this 2-qbit register, then either $|0\rangle$ or $|1\rangle$ will be obtained randomly with the corresponding probabilities of $|a|^2$ and $|b|^2$, respectively. Provided the measuring equipment, observes 0, then the measurement of the second qbit can only lead to 0. Similarly, if the first qbit is 1, the second qbit will also be 1. This is because our 2-qbit system $|\varphi\rangle$ contains the superposition of two classic basis states and the measurement is only capable of opting either for $|00\rangle$ or for $|11\rangle$. Plausibly, for $|\psi\rangle$ having four basis states any combination of 0 and 1 may be encountered. Hence it appears as if there was a mysterious connection between the two qbits and indeed, there is! Carefully designed experiments based on the Bell inequalities of [144] or on the inequalities formulated in [145] have demonstrated that this interesting effect remains valid even if the qbits of $|\varphi\rangle$ are delivered to two arbitrarily distant locations. Furthermore, surprisingly *the propagation of this linkage between the two qbits after the first measurement takes zero time.*

Let us now introduce some further related terminology. The quantum states whose decomposition exists in the above-mentioned sense are referred to as *product states*, while qbits/qregisters tied together by the above-mentioned phenomenon are referred to as *entangled states*. Entanglement is an efficient tool of quantum computing and communications, which facilitates the 'science-fiction concept of teleporting, communication over zero-capacity channels' and - more realistically - fast algorithms. However, at the time of writing commercial 'quantum PCs' and 'quantum-phones' are absent from the shelves of electronics shops. Nonetheless, emerging quantum communications applications are already available on the market, for example in the field of quantum cryptography, while a range of further applications are close to practical implementation.

Finally, it is worth mentioning that despite the fact that Einstein initiated the quest for the clarification of the quantum rules, he never accepted the concept of entanglement (he referred to it as a '*spooky action at a distance*') in his highly-acclaimed thought-experiment known as the Einstein, Podolsky and Rosen (EPR) paradox published in 1935 [146]. Just to mention one of the 'spooky actions at a distance', when observing an entangled qbit, it sheer observation instantaneously changes its entangled counterpart, regardless of its geographic position, which implies a propagation velocity higher than that of light...

### B. Quantum-assisted Communications

In wireless systems there is always a trade-off between reducing the transmit power and mitigating the resultant signal-degradation imposed by the transmit-power reduction with the aid of sophisticated receiver algorithms, when considering the total energy consumption. This is because more sophisticated receivers dissipate more power. The associated relationship becomes even more complex in a multi-user networking context. Ideally, our design objective should be that of minimizing the *total* power consumption assigned to both transmission over the ether and to the signal-processing electronics. More explicitly, it is neither economical nor 'green' to invoke the most powerful available signal processing techniques owing to their high power consumption, unless the achievable *total* transmit power reduction or some other design constraints, such a severely limited available frequency band justifies this. This is because in general, the bandwidth-efficiency and power-efficiency are inversely proportional to each other.

Quantum-assisted wireless communications exploits the extra computing power offered by quantum mechanics based architectures. This subsection summarises some recent results in quantum computing and the corresponding application areas in wireless communications.

Grover published his quantum-domain database search algorithm in [147] and [148] to illustrate its benefits. This algorithm is capable of searching through *unsorted* databases by initiating parallel data base queries and evaluating the answers in parallel. Notwithstanding its efficiency, Grover's solutions require several consecutive 'query-evaluation' pairs, nonetheless, they only necessitate on the order of $\mathcal{O}(\sqrt{N})$ iterations to carry out the search compared to the classically required $\mathcal{O}(N)$ complexity, where $N$ is the size of the search space. It was shown to be *optimal in terms of the number of iterations in [149]*. As a further advance, Boyer, Brassard, Hoyer and Tapp [150] then enhanced the original algorithm by making it capable of finding $M$ occurrences of the requested entry in the database and introduced upper bounds for the required number of operations, which was found to be on the order of $\mathcal{O}(\sqrt{N/M})$. Unfortunately the results of Grover's algorithm are probabilistic in the sense that they provide the database index of the requested item with an error probability of $P_{err} \approx M/N$. Following a range of further improvements, such as for example the elimination of the probability of error without increasing the computational complexity imposed as well as tolerating the non-uniform distribution of the input database indeces provided for example by another quantum algorithm, the most general form of quantum database search algorithms was then disseminated in [140]. Similar database search problems are also often encountered in wireless communications, for example, when finding the most suitable resource allocation, which results in the lowest level of co-channel interference across the entire network.

To elaborate a little further on the potential quantum-based communications techniques of the future, searching through unsorted databases may be viewed as being equivalent to finding certain points of a function $y = f(x)$, such as the minimum or maximum of a CF. Just to mention a few examples in wireless communications, we are typically looking for the specific $K$-user bit vector, which maximizes the CF of a MUD, or of a multi-stream MIMO-detector, etc. Unfortunately, classic MUD or MIMO-detector solutions suffer from a high computational complexity if the database is unsorted or equivalently, if the CF concerned has numerous local minimum/maximum points.

For example, the ML MUD of a $K = 10$-user 64QAM wireless system would have to evaluate the MMSE CF for all legitimate symbol combinations, namely $64^{10}$ times, which is clearly unrealistic. Furthermore, multiple CF-optima may exist, when there are more transmit antennas in a MIMO system than the number of receiver antennas, because in this scenario the channel-matrix becomes a non-square, rank-deficient and hence non-invertible matrix. Similar problems are also often encountered in resource-allocation techniques, when for example a MS has to search through the list of potential hand-over target BSs to find the one providing the best signal quality. In [151] the classic logarithmic search - which is known to be efficient for sorted data bases - has been combined with quantum-based 'existence testing' in order to answer the question, as to whether the data base does or does not contain a specific entry at all? As alluded to above, MUDs or multi-stream MIMO detectors may also be viewed as the optimization of carefully chosen CFs.

The scope of the above-mentioned $K$-user DS-CDMA MLS CF may be further extended, in order to handle an entire burst of symbols $b_k[i]$ for the $K$ users. Naturally, in the absence of any channel-induced dispersion, ie. Inter-Symbol Interference (ISI) there would be no benefit in considering several consecutive bits during the decision process, since they are independent of each other. However, in case of practical ISI-contaminated dispersive channels this so-called *jointly optimum decision* would mitigate/eliminate both the MUI and ISI by estimating $R$ symbols - rather than a single symbol - of all the $K$ users jointly during a given DS-CDMA transmission burst. This results in a potentially excessive search-space, which would be unrealistic to search through with the aid of conventional search/detection techniques. Hence classic MUDs generate an estimate $\tilde{b}_k[i]$ for the $K$ users on a symbol by symbol basis.

Having considered the basic philosophy of 'quantumised' search/optimization techniques, let us now continue by stipulating the optimization CF, which potentially has a more grave influence on the final result, than the choice of the specific optimization technique employed. We can use for example the classic Zero-Forcing (ZF) and the Minimum Mean Square Error (MMSE) CF, as well as the more recently-proposed direct Minimum Bit-Error Ratio (MBER) CF, which were used in a MIMO multi-stream detection context in [11] and the *maximum likelihood sequence* (MLS) estimation criterion.

The first version of the above-mentioned quantum-assisted multiuser detection (QMUD) method was published in [152] and improved in [140] and by Zhao *et al.* in [153]. A range of other closely related approaches have been introduced during the period of 2000 - 2010. For example, Li *et al.* applied quantum neural networks in [154], while Gao *et al.* [155] introduced a quantum bee colony optimization (QBCO) technique for solving the above-mentioned MUD problem.

### C. Quantum-based Communications

The rudiments of classic information theoretic capacity were highlighted and relied upon in the earlier parts of this treatise, but its generalization to quantum information theory [142] is beyond the scope of this paper. Instead, we adopt a more practical approach and review a few quantum communication related aspects.

Suffice to say that the classic Shannonian entropy $H(p)$ has to be replaced by the so-called quantum entropy $S(\rho)$ of von Neumann [141], when we want to quantify the information content of a quantum source. Naturally, the $\mathcal{N}$ quantum channels may be regarded as an extension of the $N$ classic channels, as detailed in [141], noting that some similarities to the complex baseband equivalent description widely used in wireless communication may be observed [156]. As an important application example, the error correction techniques of classic wireless communications have also been extended to quantum channels. More specifically, various block-coding methods were developed in [141], while pilot-symbol based solutions have appeared in [157].

Let us now turn our attention to the realms of opportunities opened up by communications over quantum channels. Naturally, their capacity is one of the key aspects of their promise in future communications. Classically the mutual information between the channel's input and output has to be maximized [158]. Naturally, in case of quantum channels the capacity had to be redefined, potentially leading to diverse scenarios to be considered. A natural distinction concerning the channel capacity definition is, whether we restrict ourselves to classic inputs/outputs or not. In the former case of classic inputs/outputs we encode the input symbols/states into quantum states, send them over the channel and make a decision at the receiver side, effectively constructing a 'classic-quantum-classic' processing chain. This is a natural approach, since humans can only process classic information. By contrast, if we do not restrict ourselves to classic inputs/outputs, we are capable of dealing with quantum channels within larger quantum systems. The most important question arising in this context is, whether quantum channels are capable at all of increasing the achievable capacity and if so, under what conditions.

The classic capacity of quantum channels has been quantified for decomposable product state inputs in form of the so-called *unentangled classical capacity*, which is also often referred to as the Holevo-Schumacher-Westmoreland (HSW) capacity denoted by $C(\mathcal{N})$. In order to highlight the power behind the application of quantum channels, let us consider a rudimentary example. The classical binary symmetric channel (BSC), which either inverts or leaves unchanged an input bit with a probability of $p = 1/2$ has zero capacity quantified as $C(N) = 1 - H(p) = 0$. However, we may readily construct an appropriate classic-single-bit to single-qbit encoding at the transmitter and the corresponding detector at the receiver [159], so that all the classic bits transmitted over the channel will be received correctly with a unity probability, i.e. we have $C(\mathcal{N}) = 1$. *As a stunning consequence, redundancy-free error correction is possible over noisy transmission media, at least for a specific subset of quantum channels.* The rationale of this extraordinary statement may be traced back to the increased degree of freedom associated with the encoding and decoding processes introduced by the $1^{st}$ Postulate.

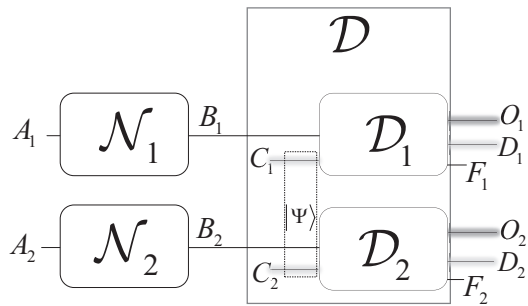

Fig. 20: Superactivation of zero-capacity quantum channels

*The science-fiction saga continues...* One of the hot research topics in this field is referred to as *superactivation* [160]. Naturally, in contrast to the previous striking example, *there are also numerous quantum channels, which have zero capacity in the context of classic information transmission.* Nonetheless, when considering two of these channels used in a parallel manner and, additionally applying the 'entanglement-controlled' decoder architecture $\mathcal{D}$ of Fig. 20, the pair of quantum channels $\mathcal{N}_1$ and $\mathcal{N}_2$ succeed in delivering classic information over the $A_1 - O_1$ and $A_2 - O_2$ links. In other words, their capacities have been (super)activated, where $A_1$ and $A2$ are the channel inputs, while $B_1$ and $B_2$ are the corresponding channel outputs linked to the decoder. This superactivation may be achieved by coupling the two subdecoders $\mathcal{D}_1$ and $\mathcal{D}_2$ with the aid of the entangled state $|\Psi\rangle$ of inputs $C_1$ and $C_2$. The outputs $O_1$ and $O_2$ of the decoder provide the payload information, while $D_i$ and $F_i$ have an auxiliary role, since they guarantee that the operator describing the overall operation of the decoder remains unitary. Current research aims to complete the set of those channels which can be superactivated.

Concerning the *entanglement assisted capacity* $C_E(\mathcal{N})$, when the qbits at the channel's input are conditioned to be entangled, a plethora of open questions are under discussion. Perhaps the most challenging one to answer is, whether entanglement is capable of increasing the attainable capacity.

A promising, but rather specific application of quantum channels is constituted by secret quantum key distribution techniques conceived for exchanging the classic encryption keys, which can be used for symmetric-key cryptography between distant locations. In order to indicate the importance of this area, information theoreticians have defined the *private capacity* $C_P(\mathcal{N})$ of quantum channels from a secure classic information transmission perspective [161].

If quantum information is fed into the quantum channel, the so-called *quantum capacity* $Q(\mathcal{N})$, which is also referred to as the Lloyd-Shor-Devetak (LSD) capacity [162] has to be considered, which is expected to be upper-bounded by the classic capacity of the same channel. In this context the MUD problem also exists in quantum channels, as discussed in [163, 164]. For detailed discussions on quantum information theory please refer to [142].

## VI. Conclusions

Many important aspects of contemporary wireless solutions have not even been touch upon owing to space-limitations, such as the benefits of Ultra-Wideband (UWB) systems [165] or Cognitive

Radios (CR) [166, 167] - just to mention a few of the essential elements of the wireless landscape.

Again, over the past three decades a 1000-fold bit rate improvement was achieved, which facilitated the introduction of the powerful new wireless services featured in the stylised illustration of Figure 2. These enticing, but bandwidth-hungry and power-thirsty multimedia services 'absorbed' the above-mentioned bitrate improvements. Furthermore, the increased popularity and wealth-creation potential of conventional mobile phones, IPhones and tablet-computers resulted in the unprecedented penetration of wireless communications, as indicated in Figure 1. As a result, the amount of tele-traffic is expected to substantially increase during the next decade, which requires new frequency bands. As a result, the RF bands gradually migrated from 450 MHz, 900 MHz, 2 GHz, 5 GHz, 60 GHz and beyond, into the THz optical domain.

Regretfully, as the carrier frequency is increased, the pathloss tends to increase and the RF propagation properties gradually become reminiscent of those of visible light. Hence the cell-size has been reduced over the past decades from the original 35 km rural GSM-cells to small urban pico-cells. Since plenty of unlicensed bandwidth is available in the optical frequency domain, cutting-edge new research is required for achieving the same level of maturity in optical wireless communications, as in RF engineering.

Additionally, radical advances are required in both quantum-information theory to set out the theoretical performance limits, like Shannon did for classic communications, and in the quantum-domain counterparts of RF transceivers capable of approaching the quantum-information theoretic predictions.

Again, not even touch upon in this treatise, nonetheless, we would like to close by alluding to flawless tele-presence, which requires the extension of the transceiver design principles provided in this treatise to compelling future services, with an ambience of joy and wonder. The recent advances in three-dimensional (3D) and holographic displays [168] facilitate immersive, flawless tele-presence with the aid of the stereoscopic video toolbox of the H.264/MPEG4 codec, provided of course that a sufficiently high-speed, high-integrity wireless link is available.

The holographic video displays available on the market at the time of writing are set to revolutionarize telecommunications, as the emergence of mobile communications did two decades ago, leading to flawless tele-presence. However, the error-resilient streaming of multiview and 3D holographic video using optical wireless and quantum-domain transceivers constitutes a further challenge...

## Acknowledgment

We would like to thank Sebastian Caban and Christian Mehlführer for their expert support of the measurements as well as Stefan Schwarz for providing LTE simulations and preparing Figures 9–11 at the University of Technology in Vienna.

We would like to thank the current and past members of the optical wireless communications groups at Oxford and Edinburgh for their contributions to the work reported in this treatise, especially Svilen Dimitrov for providing the results on optical wireless networking.

Harald Haas would also like to acknowledge the support of the work on the optical wireless networking by Airbus Germany.

Lajos Hanzo would like to thank his colleagues at the University of Southampton, UK, in particular Jos Akhtman and Raymond Steels for the enlightenment gained from past and present collaborations.

## About the Authors

**Sandor Imre** (*Member, IEEE*) was born in Budapest, Hungary, in 1969. He received the M.Sc. degree in electrical engineering, the Dr.Univ. degree in probability theory and mathematical statistics, and the Ph.D. degree in telecommunications from the Budapest University of Technology (BME), Budapest, Hungary, in 1993, 1996, and 1999, respectively, and the D.Sc. degree from the Hungarian Academy of Sciences, Budapest, Hungary, in 2007. Currently he is Head of Telecommunications Department at BME. He is also Chairman of the Telecommunication Scientific Committee of the Hungarian Academy of Sciences. Since 2005, he has been the R&D Director of the Mobile Innovation Centre. His research interests include mobile and wireless systems, quantum computing, and communications. He has made wide-ranging contributions to different wireless access technologies, mobility protocols, security and privacy, reconfigurable systems, quantum-computing based algorithms, and protocols.

Prof. Imre is on the Editorial Board of two journals: Infocommunications Journal and Hungarian Telecommunications.

**Laszlo Gyongyosi** (*Member, IEEE*) received the M.Sc. degree in computer science (with honors) from the Budapest University of Technology and Economics (BUTE), Budapest, Hungary, in 2008, where he is currently working toward the Ph.D. degree at the Department of Telecommunications. His research interests are in quantum channel capacities, quantum computation and communication, quantum cryptography, and quantum information theory. Currently, he is completing a book on advanced quantum communications, and he teaches courses in quantum computation. Mr. Gyongyosi received the 2009 Future Computing Best Paper Award on quantum information. In 2010, he was awarded the Best Paper Prize of University of Harvard, Cambridge, MA. In 2010, he obtained a Ph.D. Grant Award from University of Arizona, Tuscon. In 2011, he received the Ph.D. Candidate Scholarship at the BUTE; the Ph.D. Grant Award of Stanford University, Stanford, CA; the award of University of Southern California, Los Angeles; and the Ph.D. Grant Award of Quantum Information Processing 2012 (QIP2012), University of Montreal, Montreal, QC, Canada.